\documentclass[preprintnumbers,amsmath,11pt,amssymb,floatfix,superscriptaddress,showpacs,nofootinbib]{revtex4}
\usepackage{graphicx}
\usepackage{epsfig}
\usepackage{bm}
\usepackage{amsfonts}

\input epsf

\begin{document}

\title{\Large Correspondence between $f(G)$ Gravity and Holographic Dark Energy via Power-law Solution}

\author{Abdul Jawad}
\email{jawadab181@yahoo.com} \affiliation{Department of Mathematics,
University of the Punjab, Quaid-e-Azam Campus, Lahore-54590,
Pakistan.}

\author{Antonio Pasqua}
\email{toto.pasqua@gmail.com} \affiliation{Department of Physics,
University of Trieste, Trieste, Italy.}

\author{Surajit Chattopadhyay}
\email{surajit_2008@yahoo.co.in, surajcha@iucaa.ernet.in}
\affiliation{ Pailan College of Management and Technology, Bengal
Pailan Park, Kolkata-700 104, India.}

\date{\today}

\begin{abstract}
In this paper, we discuss cosmological application of holographic
Dark Energy (HDE) in the framework of $f(G)$ modified gravity. For
this purpose, we construct $f(G)$ model with the inclusion of HDE
and a well-known power law form of the scale factor
$a\left(t\right)$. The reconstructed $f(G)$ is found to satisfy a
sufficient condition for a realistic modified gravity model. We find
quintessence behavior of effective equation of state (EoS) parameter
$\omega_{DE}$ through energy conditions in this context. Also, we
observe that the squared speed of sound $v_s^2$ remains negative which shows
the instability of HDE $f(G)$ model.
\end{abstract}


\maketitle

\section{Introduction}

Accelerated phenomenon of the universe is now well established
thanks to many observations\cite{Spergel,Perlmu}. It is suggested in
the studies that the universe exhibits spatially flat scenario and
contains dark energy (DE) as a major component with negative
pressure and 30$\%$ of dust matter consisting of cold dark matter
(CDM) plus baryons. The contribution of radiation can be considered
practically negligible. For related review works see the references
\cite{Copeland,Tsuza}. In order to understand the DE nature, it is
required to clarify about its candidates either it is a cosmological
constant ($\Lambda$) or a form of dynamical model \cite{Tsuza}. The
dynamical DE models can be differentiated from the cosmological
constant through the tool of EoS parameter
$w_{DE}=p_{DE}/\rho_{DE}$, where the numerator and denominator
indicate, respectively, the density and the pressure of DE. Several
candidates of DE have been proposed \cite{Copeland}.

It was shown through the data analysis of SNe Ia that these
dynamical models provide more consistency with present scenario of
the universe as compare to $\Lambda$. A detailed analysis of DE can
be found in \cite{14}. The development in the study of black hole
theory and string theory results the holographic principle which
states that \textit{the number of degrees of freedom of a physical
system should be finite, it should scale with its bounding area
rather than with its volume and it should be constrained by an
infrared cut-off.}

The Holographic DE (HDE) is one of the most interesting dynamical
model and is based on the holographic principle proposed by
\cite{17}. It has also been constrained and tested by various
astronomical schemes and with the help of the anthropic principle
\cite{27}. By the inclusion of holographic principle into cosmology,
it can be found the upper bound of the entropy contained in the
universe \cite{17}. Through this bound, Li \cite{28} proposed the
constraint on the DE density:
\begin{equation}
\rho_{DE}= 3c^2 M_p^2 L^{-2}, \label{1}
\end{equation}
where $c^2,~L,~M_{p}=\left( 8\pi G \right)^{-1/2} =10^{18} GeV$
indicates a numerical constant, the IR cut-off and the reduced Planck mass,
respectively.\\
There is another form of dynamical DE models which is proposed
through large-distance modification of gravity \cite{Tsuza}.
Importance of modified gravity for late acceleration of the universe
has been reviewed by \cite{Tsuza, Nojiri1, Clifton}. In this class,
$f(R)$ gravity, $f(T)$ gravity, braneworld models, Galileon gravity,
Gauss-Bonnet gravity etc exist. All these theories of modified
gravity are thoroughly discussed in \cite{Tsuza}. The theory of
scalar-Gauss-Bonnet gravity, named as $f(G)$ has been proposed by
\cite{nojiri1}. Since the focus of the present work is $f(G)$
gravity, we explore the existing literature on $f(G)$ gravity.
Myrzalulov et al \cite{Myrza1} have found that $f(G)$ gravity
describes the accelerated expansion phenomenon of the universe and
also explain the radiation$/$matter dominated eras. In the same
paper, it was shown that $f(G)$ models have ability to reproduce DE
and inflation and can be modeled as an inhomogeneous fluid with a
dynamical EoS parameter.

In a recent work \cite{rastkar}, $f(G)$ models from the power law
solutions by coupling gravity with perfect fluid has been found. It
was observed that a special form of $f(G)$ models exist for a
phantom phase of the universe. Garcia et al \cite{Garcia} considered
specific realistic forms of $f(G)$ gravity and presented general
inequalities through energy conditions and explored the viability of
the various forms of $f(G)$ gravity through cosmographic analysis.

The main purpose of this work is to reconstruct $f(G)$ model by
using HDE and check its viability through different cosmological
parameters. For this purpose, we assume exact power law solution of
scale factor. Also we investigated the validity of energy conditions
and stability of the model. We adopt the pattern of the paper as
follows. In Section \textbf{2}, we provide the basic formalism of
$f(G)$ gravity and HDE. We reconstruct the HDE $f(G)$ model taking
into account the correspondence process. Subsection \textbf{A}
inherits the viability of energy conditions for the corresponding
model whereas stability of the model is investigated in the
subsection \textbf{B}. The results of the paper are summarized in
the last Section.

\section{Correspondence between $f(G)$ gravity and HDE}

The $f(G)$ action that describes Einstein's gravity coupled with
perfect fluid plus a Gauss-Bonnet ($G$) term is given by
\cite{nojiri1,rastkar}:
\begin{equation}
S=\int
d^4x\sqrt{-g}\left[\frac{1}{2\kappa^{2}}R+f(G)+\mathcal{L}_{m}\right], \label{2}
\end{equation}
where
$G=R^{2}-4R_{\mu\nu}R^{\mu\nu}+R_{\mu\nu\lambda\sigma}R^{\mu\nu\lambda\sigma}$
($R$ represents the Ricci scalar curvature, $R_{\mu\nu}$ represents
the Ricci curvature tensor and $R_{\mu\nu\lambda\sigma}$ represents
the Riemann curvature tensor), $\kappa^{2}=8\pi G$, $g$ is the
determinant of the metric tensor $g_{\mu \nu}$ and $\mathcal{L}_{m}$
is the Lagrangian of the matter present in the universe. The
variation of the action $S$ with respect to $g_{\mu\nu}$ generates
the field equations. In this paper, a special form of $f(G)$ gravity
proposed by \cite{rastkar} is being used. In case of flat FRW metric
(corresponding to curvature parameter $k=0$), the Ricci scalar
curvature  $R$ and the Gauss-Bonnet invariant $G$ become,
respectively, $R=6(\dot{H}+2H^{2})$ and $G=24H^{2}(\dot{H}+H^{2})$,
where the dot represents the time derivative. The first FRW equation
(with $8 \pi G=1)$ takes the form:
\begin{equation}
H^{2}=\frac{1}{3}(Gf_{G}-f(G)-24\dot{G}H^{3}f_{GG}+\rho_{m})=\frac{1}{3}(\rho_{G}+\rho_{m}),
\label{3}
\end{equation}
where $f_G=\frac{df}{dG}$ and $f_{GG}=\frac{d^2 f}{dG^2}$. Earlier,
Setare \cite{Setare1} reconstructed $f(R)$ gravity from HDE. In this
Section, we shall discuss a reconstruction of the above form of
$f(G)$ gravity in HDE framewok. The HDE density is defined as:
\cite{Setare1}
\begin{equation}
\rho_{\Lambda}=\frac{3 c^{2}}{R_{h}^{2}}, \label{4}
\end{equation}
where $R_{h}$ represents the future event
horizon which is defined as:
\begin{equation}
R_{h}=a\int_{t}^{\infty}\frac{dt}{a}=a\int_{a}^{\infty}\frac{da}{Ha^{2}}. \label{5}
\end{equation}

The dimensionless DE density is defined by the critical energy
density $\rho_{cr}=3H^{2}$ as follow:
\begin{equation}
\Omega_{\Lambda}=\frac{\rho_{\Lambda}}{\rho_{cr}}=\frac{c^{2}}{R_{h}^{2}H^{2}}. \label{6}
\end{equation}
The derivative of $R_h$ with respect to the cosmic time $t$ is given by:
\begin{equation}
\dot{R}_{h}=R_{h}H-1=\frac{c}{\sqrt{\Omega_{\Lambda}}}-1. \label{7}
\end{equation}
Using the conservation equation, the EoS parameter for HDE is
obtained by \cite{Setare1} as follow:
\begin{equation}
\omega_{\Lambda}=-\left(\frac{1}{3}+\frac{2\sqrt{\Omega_{\Lambda}}}{3c}\right). \label{8}
\end{equation}
If the universe is dominated by the HDE, i.e.,
$\Omega_{\Lambda}\rightarrow 1$, then we observe different
conditions on $c$. The $\omega_{\Lambda}$ represents the
quintessence like behavior for $c>1$ while phantom dominated
universe $(\omega_{\Lambda}<-1)$ is observed for the values of $c$
less than $1$. For $c=1$, the EoS parameter of HDE indicates the de
Sitter phase of the universe. Hence the value of $c$ is important in
the evolving universe dominated by HDE. In order to apply the
correspondence, we equate $\rho_{G}=\rho_{\Lambda}$. It follows that:
\begin{equation}
3H^{2}\Omega_{\Lambda}=Gf_{G}-f(G)-24\dot{G}H^{3}f_{GG}, \label{9}
\end{equation}
yielding:
\begin{equation}
f(G)=Gf_{G}-3 H^{2}\Omega_{\Lambda}-24\dot{G}H^{3}f_{GG}. \label{10}
\end{equation}
Here we assume the scale factor $a\left(t\right)$ in the form of
exact power law solution of the field equation as \cite{Setare1}:
\begin{equation}
a(t)=a_{0}(t_{s}-t)^{\gamma}, \label{11}
\end{equation}
where, $a_{0}$, $t_{s}$ and $\gamma$ are constants. In particular,
$a_0$ represents the present day value of the scale factor and $t_s>t$
is the future singularity finite time. This type of scale factor may
give the sudden singularity (type II) or singular curvature and
$\dot{H}$ (type IV) for positive $\gamma$. With this choice of scale
factor, the Hubble parameter $H$ and $G$ become, respectively,
$H=-\frac{\gamma}{t_{s}-t}$ and
$G=\frac{24(\gamma-1)\gamma^{3}}{(t_{s}-t)^{4}}$. Inserting
Eq. (\ref{11}) in Eq. (\ref{10}), we can rewrite the differential
equation given in Eq. (\ref{10}) as follow:
\begin{equation}
\frac{2304(\gamma-1)\gamma^{6}}{(t_{s}-t)^{8}}f_{GG}+\frac{24
\gamma^{3}(\gamma-1)}{(t_{s}-t)^{4}}f_{G}-f(G)=\frac{3\gamma^{2}-3c^{2}(\gamma-1)^{2}}{(t_{s}-t)^{2}}.
\label{12}
\end{equation}
In terms of $G$, Eq. (\ref{12}) can be rewritten as:
\begin{equation}
\frac{4}{\gamma-1}G^{\frac{3}{2}}f_{GG}+\sqrt{G}f_{G}-G^{-\frac{1}{2}}f(G)=
\frac{3\gamma^{2}-3c^{2}(\gamma-1)^{2}}{\sqrt{24\gamma^{3}(\gamma-1)}},
\label{14}
\end{equation}
which is a differential equation of order two which solution is
given by:
\begin{equation}
f(G)=-\frac{8}{\gamma+1}\sqrt{G}+G\left(G^{-\frac{3+\gamma}{4}}C_{1}+C_{2}\right),
\label{15}
\end{equation}
where $C_1$ and $C_2$ are integration constants. This is the $f(G)$
model in the HDE scenario with power-law form of scale factor.
\begin{figure}[h]
\begin{minipage}{14pc}
\includegraphics[width=16pc]{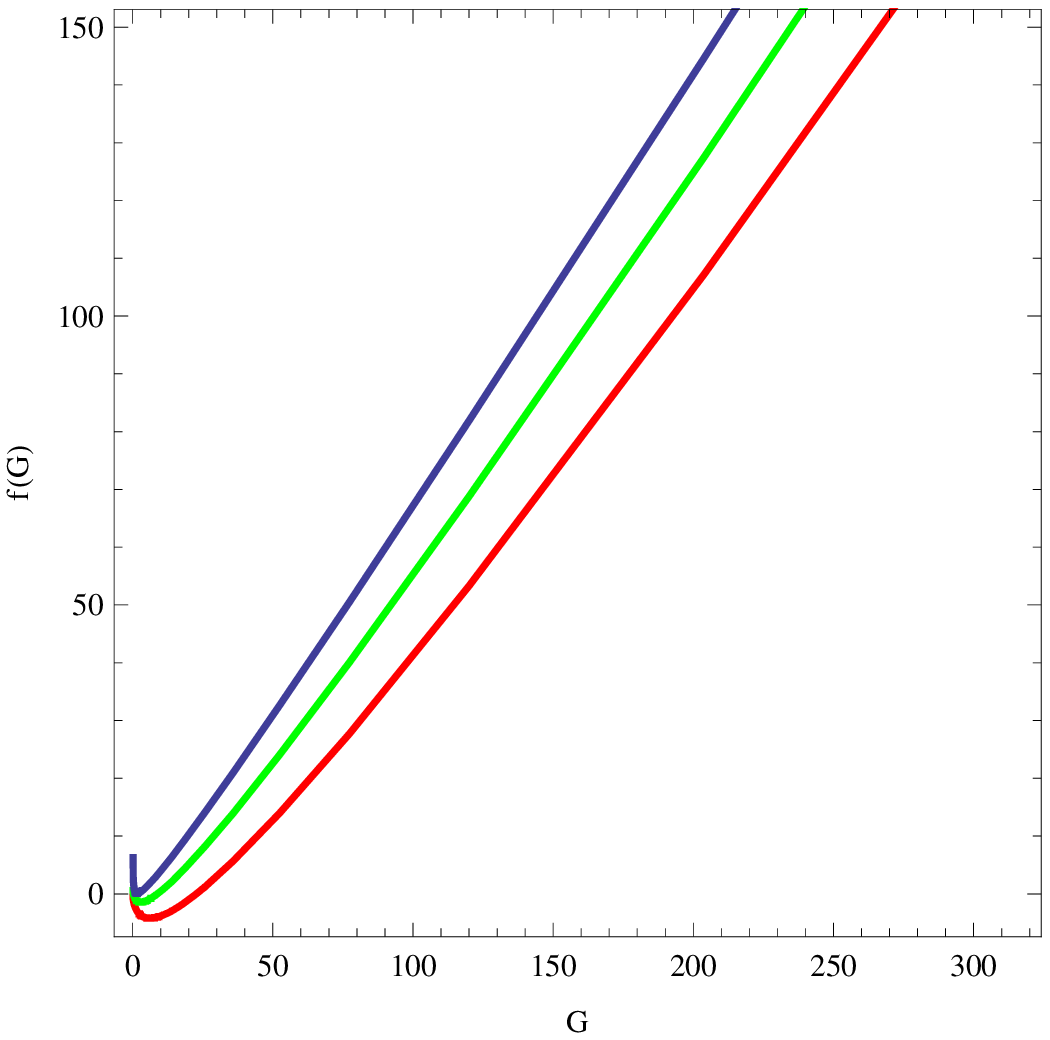}
\caption{\label{label}Evolution of HDE $f(G)$ model versus $G$ with
power-law scale factor. Red, Green and Blue lines correspond,
respectively, to $\gamma=1.02,~2.2~\textrm{and}~5.2$.}
\end{minipage}\hspace{3pc}%
\begin{minipage}{14pc}
\includegraphics[width=16pc]{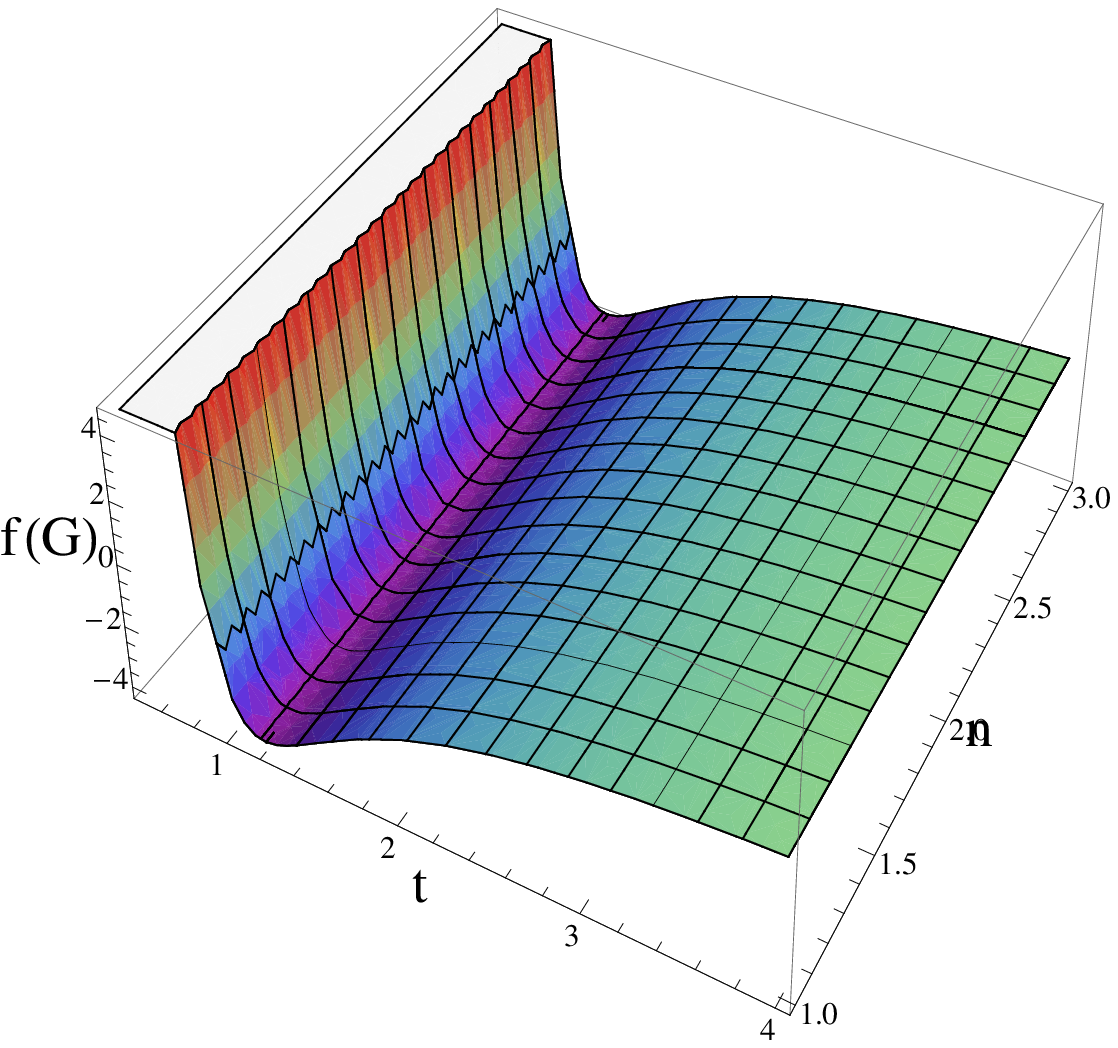}
\caption{\label{label}Evolution of HDE $f(G)$ model versus $t$ for a
range of values of $\gamma$.}
\end{minipage}\hspace{3pc}%
\end{figure}
We plot $f(G)$ model against $G$ as shown in Figure \textbf{1}. We
find that $f(G)$ is exhibiting increasing pattern with increasing
$G$. Furthermore, we observe that
\begin{equation}
\lim_{G\rightarrow 0}f(G)=0. \label{16}
\end{equation}
This is a sufficient condition to ensure that HDE $f(G)$ model is a
realistic model \cite{rastkar}. Hence, the reconstructed $f(G)$
given in Eq. (\ref{15}) is a consistent modified gravity with HDE in
flat space. The $f(G)$ model can be interpreted as a function of
cosmic time in the following way:
\begin{eqnarray}\nonumber
f(G)&=&-\frac{16\sqrt{6}}{\gamma+1}\sqrt{\frac{(\gamma-1)\gamma^{3}}{(t_{s}-t)^{4}}}
\\\label{c}&+&
\frac{24\gamma^{3}(\gamma-1)}{(t_{s}-t)^{4}}\left[C_{2}+C_{1}2^{-\frac{3(3+\gamma)}{4}}3^{-\frac{\gamma+3}{4}}
\left(\frac{(\gamma-1)\gamma^{3}}{(t_{s}-t)^{4}}\right)^{-\frac{3+\gamma}{4}}\right].
\end{eqnarray}
Figure 2 represents the graph of HDE $f(G)$ model versus $t$ for a
range of values of $\gamma>1$. Initially, it shows a decreasing
behavior and it assumes negative values. After a short interval, we
find the increment in $f(G)$ and it becomes positive. Hence, the
model given in Eq. (\ref{c}) describes the decaying process firstly and then
starts increasing, becomes positive during the evolution of the
universe.

In the subsequent Section we shall consider the energy conditions
and stability of the reconstructed  HDE $f(G)$ gravity model.

\subsection{Energy conditions for HDE $f(G)$ gravity}

We now find the energy conditions for reconstructed $f(G)$ gravity
in HDE scenario. In context of modified theories of gravity, the
phenomenon of energy conditions have been widely studied \cite{C}.
Its various forms, namely null energy condition (NEC), weak energy
condition (WEC), dominant energy condition (DEC) and strong energy
condition (SEC) are found through attractiveness of gravity along
with Raychaudhuri equation \cite{Garcia,C}. The expansion sceanrio
of the universe \cite{E}, phantom field potential \cite{F} and
evolution of the deceleration parameter $q$ \cite{G} have been
explored using these conditions in general relativity. Sadeghi et
al. \cite{H} and Garcia et al. \cite{Garcia} investigated the
viability of some specific forms of $f(G)$ model by using
approximated current values of Hubble, jerk, deceleration and snap
parameters with the help of these conditions. They also explored
that the late-time de-Sitter solution is stable and the investigated
the existence of standard radiation/matter dominated eras through
WEC and SEC. Using effective approach, the energy conditions are
given, respectively, by \cite{Garcia}:
\begin{itemize}
    \item NEC: $\rho_{eff}+p_{eff}\geq 0$,
    \item WEC: $\rho_{eff}\geq 0,\quad \rho_{eff}+p_{eff}\geq 0$,
    \item DEC: $\rho_{eff}\geq 0,\quad \rho_{eff}\pm p_{eff}\geq 0$,
    \item SEC: $\rho_{eff}+3p_{eff}\geq 0,\quad \rho_{eff}+p_{eff}\geq 0$.
\end{itemize}
\begin{figure}
\begin{minipage}{14pc}
\includegraphics[width=16pc]{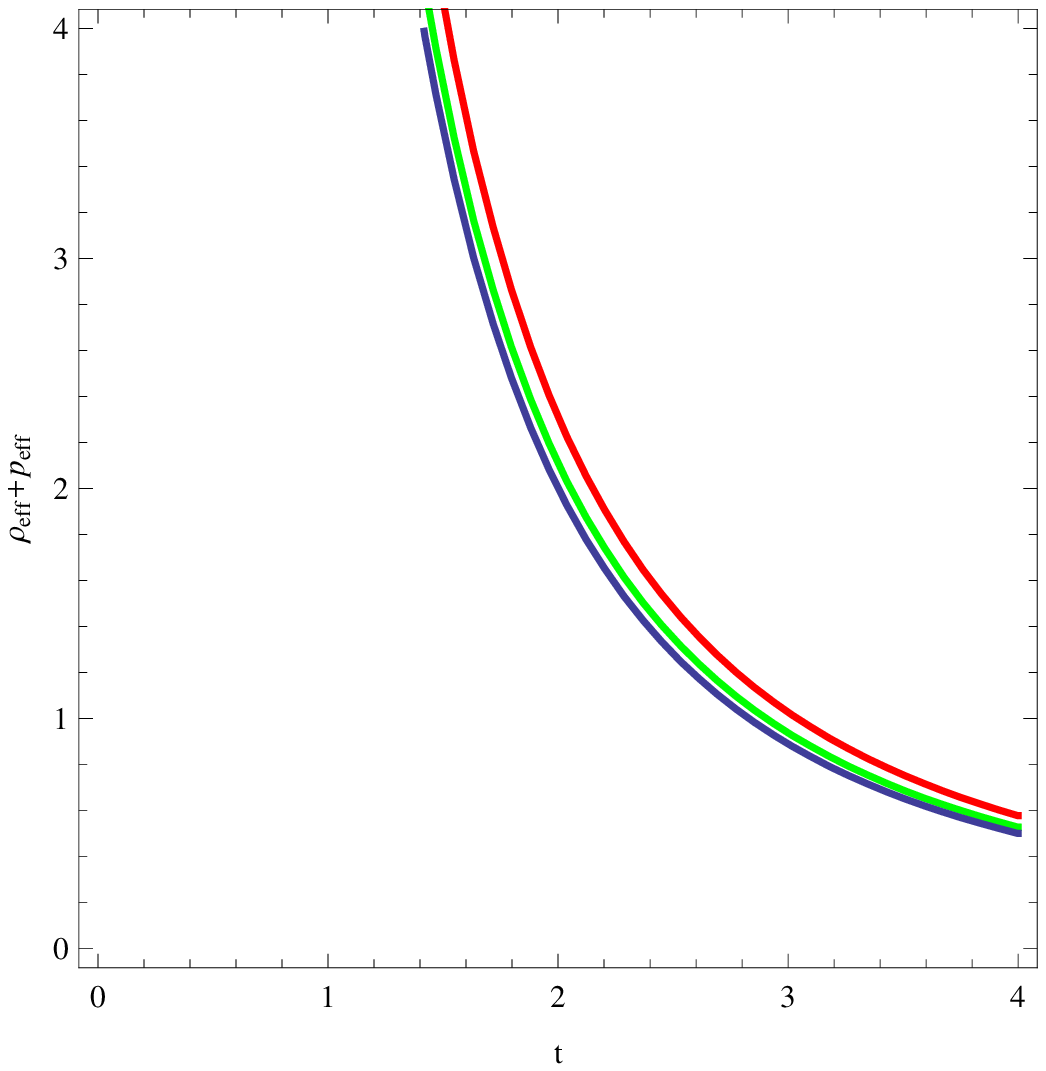}
\caption{\label{label}Behaviour of $\rho_{eff}+p_{eff}$ versus $t$
for different values of $\gamma>1$.}
\end{minipage}\hspace{3pc}%
\begin{minipage}{14pc}
\includegraphics[width=16pc]{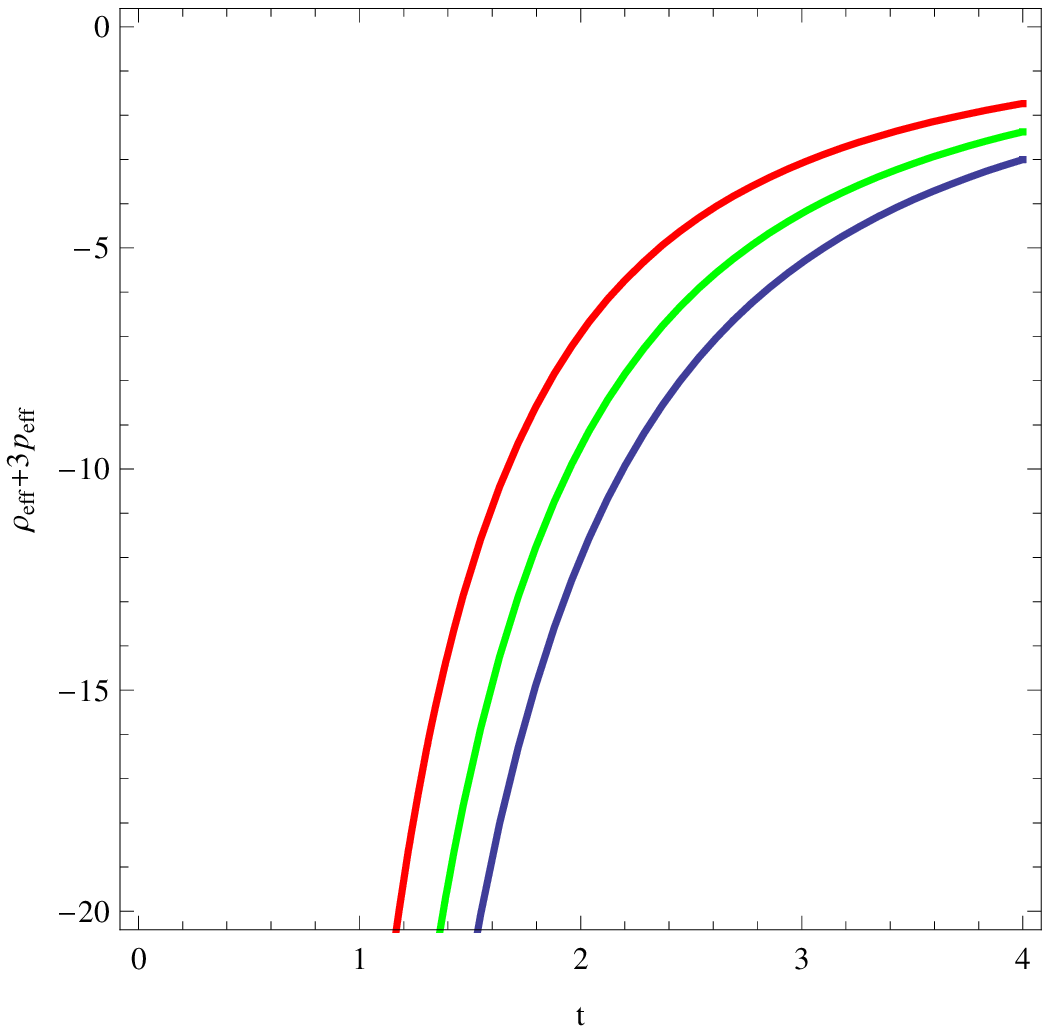}
\caption{\label{label}Behaviour of $\rho_{eff}+3p_{eff}$ versus $t$
for different values of $\gamma>1$.}
\end{minipage}\hspace{3pc}%
\end{figure}

The NEC and the WEC are simple and important in the sense that the
violation of NEC results to the violation of remaining energy
conditions. It guarantees the validity of second law of
thermodynamics and it represents the reduction of energy density
with expansion of the universe. However, the violation of SEC
represents the accelerated expansion of the universe. For HDE $f(G)$
gravity, the effective energy density and the effective pressure are
obtained as
\begin{eqnarray}\nonumber
\rho_{eff}&=&\rho_{m}+\frac{1}{2}\left\{G(C_{2}+C_{1}G^{-\frac{3+\gamma}{4}})+
6H^{2}+24H^{3}f(G)\dot{G}-\frac{8\sqrt{G}}{1+\gamma}\right.\\\label{d}&-&
\left. G^{2}\left[G(C_{2}+ C_{1}G^{-\frac{3+\gamma}{4}})-\frac{8\sqrt{G}}{\gamma+1}\right]\right\},\\
p_{eff}&=&p_{m}+\frac{1}{2}\left\{G(C_{2}+C_{1}G^{-\frac{3+\gamma}{4}})-
\frac{8\sqrt{G}}{\gamma+1}+4608H^{4}(H\ddot{H}+4H^{2}\dot{H}+2\dot{H}^{2})^{2}\right.\nonumber
\\ &\times&
\left.\left[\frac{1}{16}C_{1}G^{-\frac{11+\gamma}{4}}(\gamma+3)(\gamma+7)-
\frac{1}{64}C_{1}G^{-\frac{11+\gamma}{4}}(\gamma+7)(\gamma+3)^{2}-\frac{3}{G^{\frac{5}{2}}(1+\gamma)}\right]\right.\nonumber
\\&+&
\left.192H^{2}(H^{2}\dddot{H}+6H^{3}\ddot{H}+8H^{4}\dot{H}+8\dot{H}\ddot{H}
+24H^{2}\dot{H}^{2}+6\dot{H}^{3})\left[\frac{C_{1}(3+\gamma)(\gamma-1)G^{-
\frac{7+\gamma}{16}}}{4}\right.\right.\nonumber \\
\label{b}&+&\left.
\left.\frac{2}{G^{\frac{3}{2}}(\gamma+1)}\right]-24H^{2}(\dot{H}+H^{2})\left[C_{2}-\frac{\gamma-1}{4}C_{2}G^{-\frac{3+\gamma}{4}}
-\frac{4}{\sqrt{G}(\gamma+1)}\right]\right\}.
\end{eqnarray}

The plot of NEC for HDE $f(G)$ model versus cosmic time $t$ taking
different values of $\gamma$ is shown in Figure \textbf{3}. We see
that for all values of $\gamma$,
$\rho_{eff}+p_{eff}>0\Rightarrow~w_{eff}>-1$. This indicates the
validity of NEC as well as WEC. Figure \textbf{4} represents the
plot of SEC versus $t$ for HDE $f(G)$ model for the same values of
$\gamma$. Initially, $\rho_{eff}+3p_{eff}$ describes the increasing
but negative behavior. With the passage of time, the model shows the
flatness in the curves for all values of $\gamma$ and inherits
negative behavior of $\rho_{eff}+3p_{eff}$. It gives the violation
of SEC as $\rho_{eff}+3p_{eff}<0\Rightarrow w_{eff}<-\frac{1}{3}$
and it ensures the accelerated expansion of the universe. Thus,
combining both conditions on the effective EoS parameter, we have
$-1<w_{eff}<-\frac{1}{3}$. This indicates quintessence-like behavior
of the EoS parameter for HDE $f(G)$ model.

\subsection{Stability of reconstructed $f(G)$ gravity}

We now consider an important quantity to check the stability of HDE $f(G)$ model, namely the squared speed of sound
$v_{s}^{2}$, defined as:
\begin{eqnarray}
v_{s}^{2}=\frac{\dot{p}_{\Lambda}}{\dot{\rho}_{\Lambda}}
\end{eqnarray}
The sign of $v_{s}^{2}$ is crucial for stability of a background
evolution. A negative value implies a classical instability of a
given perturbation in general relativity \cite{myung,kim}. Myung
\cite{myung} has observed that $v_{s}^{2}$ for HDE is always
negative for the future event horizon as IR cutoff, while for
Chaplygin gas and tachyon, it is observed non-negative. Kim et al.
\cite{kim} found that the squared speed of sound for agegraphic DE
is always negative leading to the instability of the perfect fluid
for the model. Also, it was found that the ghost QCD \cite{20g} DE
model is unstable. Recently, Sharif and Jawad \cite{sharif} have
shown that interacting new HDE is characterized by negative
$v_{s}^{2}$.

We plot $v_{s}^{2}$ versus $t$ taking the
power-law scale factor in the HDE $f(G)$ model as shown in Figure
\textbf{5}. Here, we consider $v_{s}^{2}$ as a
ratio of the effective pressure and energy densities given, respectively, in Eqs. (17) and
(18), i.e. $v_{s}^{2}=\frac{\dot{p}_{eff}}{\dot{\rho}_{eff}}$. We
observe that it remains negative for the present and future epoch.
This shows that $f(G)$ model in HDE scenario with power-law scale
factor is classically unstable.
\begin{figure}
\begin{minipage}{14pc}
\includegraphics[width=16pc]{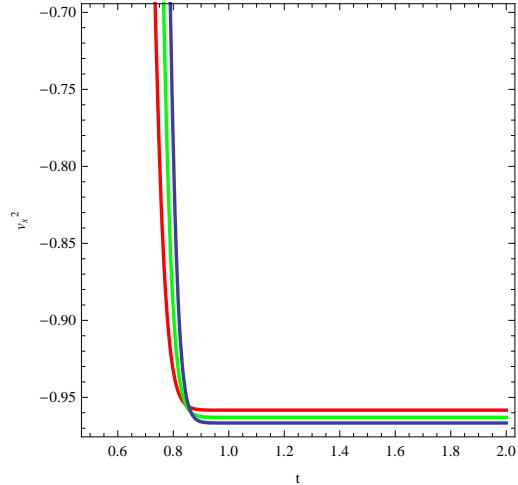}
\caption{\label{label}Behaviour of $v_{s}^{2}$ for different values
of $\gamma>1$.}
\end{minipage}\hspace{3pc}%
\end{figure}

\section{Conclusions}
Many different candidates have been proposed as possible candidates
of DE. Modified gravity has recently emerged as a possible unification of DE
and DM. This work aims at a cosmological application of HDE density
in the framework of a modified gravity, named as $f(G)$ gravity. In
this framework, we have considered the EoS parameter $\omega_{DE}$ of the HDE
density. We have developed a correspondence between the energy
densities of $f(G)$ gravity and HDE and obtained a second order
differential equation which was solved for
$f(G)$ with the help of power-law form of scale factor $a$. The validity
of NEC and SEC is investigated for this model and stability is
checked through squared speed of sound $v_{s}^{2}$.

We have found that the obtained model given in  Eq. (\ref{15}) is a
consistent modified gravity model with HDE in flat space. It gives
an increasing behavior with the passage of time. The NEC as well as
the WEC are always hold while SEC is violated for this model. The
violation of SEC gives $-1<w_{eff}<-\frac{1}{3}$ which shows the
quintessence epoch of the accelerated expansion of the universe.
Also the reconstructed HDE $f(G)$ model always remain instable like
HDE with event horizon as IR cutoff and interacting new HDE models.

\subsection{Acknowledgements}
One of the authors (SC) sincerely acknowledges the facilities
provided to him by the Inter-University Centre for Astronomy and
Astrophysics (IUCAA), Pune, India, during his visit in November,
2012 under the Visitor Associateship Programme. Financial support
from DST, Govt. of India under project SR/FTP/PS-167/2011 is duely
acknowledged by SC.\\

\end{document}